\documentclass[11pt,letterpaper]{article}

\usepackage[T1]{fontenc}
\usepackage{amsmath,amssymb,mathtools,amsthm}
\usepackage{libertine}
\usepackage[libertine]{newtxmath}
\usepackage[scaled=.92]{inconsolata}
\usepackage[margin=1in]{geometry}
\usepackage{microtype}
\usepackage{graphicx}
\usepackage{booktabs,tabularx,array}
\usepackage{float}
\usepackage[font=small,labelfont=bf]{caption}
\usepackage[round,authoryear]{natbib}
\usepackage{xcolor}
\usepackage{aliascnt}
\usepackage{hyperref}
\usepackage[nameinlink,capitalize,noabbrev]{cleveref}
\usepackage{authblk}

\definecolor{linkblue}{RGB}{35,70,115}
\hypersetup{
  colorlinks=true,
  linkcolor=linkblue,
  citecolor=linkblue,
  urlcolor=linkblue,
  pdftitle={Collateral Architecture and Correlated Slashing in Liquid Restaking}
}
\theoremstyle{plain}
\newtheorem{theorem}{Theorem}[section]
\newaliascnt{proposition}{theorem}
\newtheorem{proposition}[proposition]{Proposition}
\aliascntresetthe{proposition}
\newaliascnt{corollary}{theorem}
\newtheorem{corollary}[corollary]{Corollary}
\aliascntresetthe{corollary}
\newaliascnt{lemma}{theorem}

\aliascntresetthe{lemma}
\theoremstyle{definition}
\newaliascnt{definition}{theorem}

\aliascntresetthe{definition}
\theoremstyle{remark}
\newaliascnt{remark}{theorem}

\aliascntresetthe{remark}

\crefname{theorem}{theorem}{theorems}
\crefname{proposition}{proposition}{propositions}
\crefname{corollary}{corollary}{corollaries}
\crefname{lemma}{lemma}{lemmas}
\crefname{definition}{definition}{definitions}
\crefname{section}{section}{sections}
\crefname{figure}{figure}{figures}
\crefname{table}{table}{tables}

\newcommand{\E}{\mathbb{E}}
\newcommand{\Prob}{\mathbb{P}}
\newcommand{\R}{\mathbb{R}}
\newcommand{\1}{\mathbf{1}}
\newcommand{\Var}{\operatorname{Var}}
\newcommand{\Cov}{\operatorname{Cov}}
\newcommand{\diag}{\operatorname{diag}}
\newcommand{\trans}{\mathsf{T}}
\newcommand{\norm}[1]{\lVert #1\rVert}
\newcommand{\cX}{\mathcal{X}}
\newcommand{\cF}{\mathcal{F}}

\numberwithin{equation}{section}

\title{Collateral Architecture and Correlated Slashing in Liquid Restaking}

\author{%
  Zhejun Zhang\textsuperscript{1} \quad
  Justin Cho\textsuperscript{1} \quad
  Francisco Marmolejo-Cossío\textsuperscript{1,2}\\
  \normalsize \textsuperscript{1}Boston College \quad
  \textsuperscript{2}Input Output Global (IOG)\\
  \small\texttt{\{zhangdmn,chojus,marmolf\}@bc.edu}
}

\date{}

\begin{document}

\maketitle

\begin{abstract}
Liquid restaking protocols allow pooled stake to support multiple slashable commitments, creating an allocation problem in which additional rewards must be balanced against correlated slash claims.  This risk depends not only on dependence between slash events, but also on how collateral is allocated or reused across commitments.  We develop a unified model ranging from \emph{unique stake}, where capital is divided among positions, to shared collateral, where the same capital supports multiple commitments.  We show that portfolio variance is governed jointly by gross exposure, exposure concentration, and failure dependence.  Spreading fixed capital reduces risk toward a correlation-dependent floor, whereas repeatedly pledging the same collateral can make risk grow quadratically with breadth.

We then study risk-sensitive allocators competing for fixed reward pools distributed proportionally to committed exposure.  A single allocator has a unique optimum, and in a symmetric benchmark greater risk sensitivity encourages diversification under unique stake but discourages additional commitments under shared collateral.  With multiple allocators, positive background stake guarantees a unique pure Nash equilibrium and supports a globally convergent projected method.

Reproducible synthetic stress tests support these mechanisms.  In the strongest common-factor environment, relaxing the reuse cap raises equilibrium reward by $2.47\times$ but aggregate $99\%$ conditional value at risk by $3.77\times$.  Competition creates overexposure relative to a planner, while ignoring covariance raises true variance by $6.3\%$.  The results provide a unified framework for collateral reuse, correlated operational risk, and incentive design in decentralized security markets.
\end{abstract}

\begin{center}
\end{center}

\section{Introduction}
\label{sec:introduction}

Proof-of-stake blockchains secure themselves by requiring validators to commit capital, known as \emph{stake}, that can be confiscated if they violate protocol rules.  This penalty, known as \emph{slashing}, creates an economic incentive to behave correctly.  Restaking extends the idea by allowing already-staked capital to support additional decentralized services.  These services rely on \emph{operators}, participants that perform validation or other tasks using delegated stake as collateral.  Operators receive rewards for providing these services, while their collateral may be slashed when specified failures occur.

Liquid-restaking protocols act as intermediaries between capital holders and these opportunities. They combine deposits into pools of capital and manage each pool's exposure across operators and services. We model each such pool as an \emph{allocator}. We use a \emph{slashable position} to mean a particular collateral-backed commitment subject to a specified slashing condition. An allocator therefore faces a portfolio problem: how much exposure should it take, and how should that exposure be distributed across positions?

The key difference from a conventional fixed-budget portfolio is that commitments need not partition capital.  Under \emph{unique stake}, capital is divided among positions, so gross slashable exposure cannot exceed the capital available.  Under \emph{shared collateral}, the same capital can support several commitments, and gross exposure can exceed capital.  EigenLayer's Unique Stake design is a concrete example of the first architecture~\citep{elip002,eigenlayerUniqueStake2026,eigenlayerMagnitudes2026}; elastic or shared-collateral models provide the opposite endpoint~\citep{durvasula2025,barzur2025}.

A simple example illustrates the distinction.  Suppose a pool contains one unit of capital and considers three slashable positions.  If the unit is divided equally, each position carries exposure $1/3$ and total gross exposure remains one.  If the same unit supports all three positions, each position carries exposure one and total gross exposure is three.  Three simultaneous slash events then generate three gross claims against one unit of capital.  Correlation governs how likely those claims are to arrive together; collateral architecture governs how many claims the unit of capital creates in the first place.

Slashing risk also depends on whether slash events tend to occur together.  Positions may share operators, software clients, cloud providers, bridges, or governance processes.  Adding a position may therefore diversify a fixed exposure budget, or it may create another claim on collateral already committed elsewhere.  A useful risk model must separate these two channels.

The problem becomes strategic when several liquid-restaking pools pursue the same reward opportunities.  If a position distributes a fixed reward pool proportionally to exposure, one allocator's commitment changes the rewards captured by others.  We ask how collateral architecture and correlated slashing shape portfolio choice, equilibrium exposure, and system-level risk.

\paragraph{Restaking as a market for economic security.}
Viewed as a market, decentralized services demand economic security and liquid-restaking pools supply it.  Reward rules determine how the available payments are divided among suppliers, while collateral rules determine how many slashable commitments a unit of capital can support.  Common operational dependencies then determine whether the resulting claims are likely to arrive together.  The economic outcome therefore depends jointly on market competition, collateral architecture, and failure dependence.

\paragraph{Approach and results.}
Each allocator controls capital $W_i$ and chooses exposure across slashable positions under a reuse cap $\kappa_i$, which limits total gross exposure to $\kappa_i W_i$.  We refer to the case $\kappa_i=1$ as \emph{unique stake}.  At the opposite extreme, setting $\kappa_i$ equal to the number of positions permits \emph{full reuse}, meaning that the entire capital pool can back every position.  Varying $\kappa_i$ between these endpoints traces what we call the \emph{collateral-reuse envelope}.  Positions offer proportional rewards and generate correlated Bernoulli slash events.  Allocators maximize reward net of expected gross claims and a variance penalty.  The model includes a latent-factor construction for coherent dependence and a planner benchmark for system-level comparisons.

Our first result shows that portfolio risk depends jointly on exposure scale and concentration.  For a portfolio $x$ of exposures across positions, let $L(x)$ denote the resulting random total gross slash claim.  In a symmetric benchmark with common slash fraction $\alpha$, marginal slash probability $p$, and pairwise slash-event correlation $\rho$,
\begin{equation}
\label{eq:intro-decomposition}
\Var(L(x))
=
\alpha^2p(1-p)G^2\left[\rho+(1-\rho)H\right].
\end{equation}
Here $G$ is total gross exposure, while $H$ measures exposure concentration, with lower values indicating that exposure is spread more evenly.  The term governed by $\rho$ captures common-shock risk and remains even as exposure is spread broadly; the term governed by $(1-\rho)H$ captures position-specific risk and declines as exposure becomes less concentrated.  Under unique stake, spreading fixed capital leaves $G$ unchanged but lowers $H$, so variance falls toward a correlation-dependent floor.  Under full reuse, adding a position increases $G$ because the same capital supports another full-pool claim.  When $\rho>0$, the common-shock component therefore grows quadratically with the number of commitments.

Second, the individual allocation problem is strictly concave.  In a symmetric breadth benchmark, risk sensitivity has opposite effects across architectures: it favors broader diversification when capital is partitioned, but fewer commitments when breadth increases gross exposure.  Third, the multi-allocator game has a unique pure Nash equilibrium when every position has positive background exposure, even though the game generally lacks an exact potential.  Strong monotonicity also gives a projected pseudo-gradient method that converges globally at a linear rate. Finally, controlled stress tests quantify how rewards and tail risk change as collateral reuse expands, the reward-capture externality, and the cost of ignoring common factors.

\paragraph{Contributions.}
The paper makes four contributions.  First, it separates underlying capital, gross exposure, and exposure concentration in a single architecture model.  Second, it derives closed-form risk and breadth results that reveal an architecture-dependent reversal: breadth provides diversification under unique stake but acts like leverage under full reuse.  Third, it establishes a unique equilibrium for heterogeneous allocators and connects strong monotonicity to convergence, sensitivity, and extensions to general convex risk costs.  Fourth, it provides a reproducible computational platform, built around a projected pseudo-gradient implementation, that evaluates architecture, competition, dependence misspecification, solver behavior, and alternative risk objectives.

The remainder of the paper proceeds from primitives to implications.  \Cref{sec:context} relates the model to current protocol designs and the surrounding literature.  \Cref{sec:model} defines exposure, correlated gross claims, rewards, and system-level risk.  \Cref{sec:risk} isolates the role of collateral architecture, after which \Cref{sec:individual,sec:equilibrium,sec:planner} study private choice, strategic competition, and planner benchmarks.  \Cref{sec:experiments} evaluates the mechanisms computationally, and \Cref{sec:discussion} discusses measurement, protocol design, and scope.

\section{Protocol Context and Related Work}
\label{sec:context}

Our analysis connects three bodies of work that are often treated separately: protocol architectures that map capital into slashable commitments, models of common operational dependence, and proportional-reward games among competing suppliers of stake.

\subsection{Restaking roles and slashable positions}

Liquid restaking links four distinct roles: capital providers, pooled intermediaries, infrastructure operators, and services that purchase slashable security.  A \emph{restaker} supplies capital.  A \emph{liquid-restaking protocol} pools deposits from restakers, delegates the pooled capital to operators, and issues depositors a liquid claim on the pool.  An \emph{operator} receives delegated stake and performs tasks for one or more \emph{restaking services}, called \emph{autonomous verifiable services} (AVSs) in EigenLayer's current terminology.  A service specifies its tasks, rewards, and slashing conditions and may organize participating operators into \emph{operator sets} for different tasks or security requirements.  In our model, an \emph{allocator} represents a liquid-restaking pool, or another strategic manager of pooled capital, rather than an individual restaker, operator, or service.

The unit of exposure in our model is not necessarily an operator or service as a whole, but a particular collateral-backed commitment.  We call such a commitment, with its own reward rule and slash event, a \emph{slashable position}.  In EigenLayer, a position can correspond to stake delegated to a particular operator and allocated through that operator to a particular operator set.  A service may create several operator sets, and the same operator may participate in several sets, so services, operators, and slashable positions are not interchangeable.

EigenLayer's current Unique Stake design makes this distinction operational.  For each asset strategy, an operator assigns \emph{magnitudes}---fractions of its delegated stake---to operator sets, and the total assigned magnitude cannot exceed the operator's available magnitude.  Stake assigned to one set is exclusively slashable by the AVS that created that set, so two sets cannot slash the same allocated unit~\citep{eigenlayerMagnitudes2026,eigenlayerUniqueStake2026}.  Earlier shared-collateral designs, in which the same delegated stake could back every service an operator joined, motivate the opposite endpoint~\citep{elip002}.  Our reuse cap spans these architectures without tying the analysis to one implementation.

A liquid-restaking protocol need not control every on-chain operator magnitude directly.  It can nevertheless select operators, contract over allocation policies, and rebalance pooled deposits.  We therefore model the economic exposures induced by those choices rather than every underlying protocol action.  The operator map in \Cref{prop:architecture-map} shows how a concrete delegation architecture induces the reduced-form reuse cap used in the analysis.

Proof-of-stake reward design has also examined how incentives shape stake-pool formation and delegation~\citep{brunjes2020reward}, and how payment schemes trade off participation, decentralization, and expenditure~\citep{kiayias2024participation}.  We take participation and delegated capital as given and study the complementary problem of allocating that capital across correlated slashable positions under alternative collateral architectures.

\subsection{Restaking security and correlated operational risk}

The EigenLayer whitepaper presents restaking as pooled cryptoeconomic security for new services~\citep{eigenlayer2023}.  Existing theory largely asks whether stake and rewards deter strategic attacks.  \citet{durvasula2025} study robustness after exogenous stake loss in a restaking network; \citet{chitra2024} model services, operators, attackers, and reward selection; \citet{barzur2025} analyze elastic allocations that may exceed validator stake; and \citet{deb2024,nag2026} study safety mechanisms and alternative shared-security architectures.  Those papers emphasize attack profitability, worst-case coalitions, and cascading security.

Our slash indicators instead represent stochastic operational or task-level events.  Positions may fail together because they share an operator, software client, cloud provider, bridge, or governance process.  The two perspectives are complementary.  An allocation can deter economically rational attacks while remaining vulnerable to common software faults; conversely, a portfolio with low operational correlation can remain insecure against a profitable coalition. Our contribution is to isolate the portfolio and market consequences of this operational-risk channel while taking rewards and marginal slash probabilities as primitives.

Surveys of liquid staking and restaking emphasize the smart-contract, operator, and interconnection risks inherited by liquid tokens~\citep{gogol2024}.  Recent empirical work studies revenue drivers and cross-protocol links~\citep{sevim2026}.  We add a formal model for the joint slash distribution of the positions backing a liquid-restaking pool.  The variance decomposition is related to classical portfolio theory~\citep{markowitz1952}, but reuse changes the budget accounting: gross exposure need not equal capital.  The Gaussian threshold construction follows latent-factor credit-risk models~\citep{vasicek2002,plackett1954} and guarantees jointly attainable Bernoulli covariances.

\subsection{Proportional sharing as coupled contests}

We assume position $j$ distributes a fixed reward pool $R_j$ proportionally to exposure.  With total strategic exposure $X_j$ and background exposure $b_j$, allocator $i$ receives $R_jx_{ij}/(b_j+X_j)$. A single position is therefore a concave pro-rata game, or equivalently a Tullock contest with unit exponent and an exogenous participant of size $b_j$.  Proportional sharing has a long connection to network pricing~\citep{kelly1998}; \citet{johnson2023} establish uniqueness for a broad one-pool pro-rata class and observe rapid best-response convergence numerically.

Our market is not a collection of independent contests.  The reuse cap links positions through each allocator's capital, and the covariance cost is generally nonseparable: the marginal value of position $j$ depends on exposure to every position correlated with it.  We analyze the resulting multi-position game using Rosen's concave-game framework~\citep{rosen1965} and strongly monotone variational inequalities~\citep{facchinei2003}. The game is generally not an exact potential game, but its proportional-reward structure still supplies enough uniform curvature for equilibrium uniqueness and global convergence of a projected pseudo-gradient method.

\section{Model}
\label{sec:model}

\subsection{Capital, exposure, and collateral reuse}

There are $n$ strategic allocators and $m$ slashable positions.  Allocator $i$ controls capital $W_i>0$ and chooses $x_i=(x_{i1},\ldots,x_{im})\in\R_+^m$, where $x_{ij}$ is its exposure to position $j$.  A reuse cap $\kappa_i\in[1,m]$ defines
\begin{equation}
\label{eq:feasible}
\cX_i(\kappa_i)
=
\left\{x_i\ge0:
 x_{ij}\le W_i\ \forall j,
 \quad \1^{\trans}x_i\le\kappa_iW_i
\right\}.
\end{equation}
The two constraints play different roles.  The coordinate bounds prevent any single position from receiving more than the allocator's full capital pool, while the aggregate constraint limits the total amount of slashable exposure created from that pool.  Thus $\kappa_i$ is not a diversification parameter but an upper bound on collateral multiplicity.  As $\kappa_i$ ranges from $1$ to $m$, the nested family $\{\cX_i(\kappa_i)\}_{\kappa_i\in[1,m]}$ forms the allocator's \emph{collateral-reuse envelope}.  Its endpoints are unique stake at $\kappa_i=1$ and full reuse at $\kappa_i=m$; intermediate values represent bounded reuse, while the distribution of exposure within the permitted set remains a separate portfolio choice.

\begin{figure}[t]
\centering
\includegraphics[width=.64\linewidth]{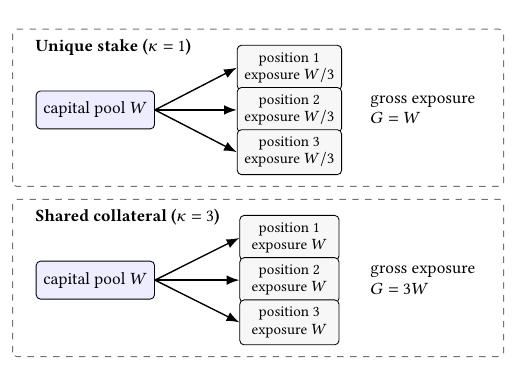}
\caption{The same capital pool and nominal breadth create different gross exposure.  Unique stake partitions capital; shared collateral creates multiple claims on the same pool.}
\label{fig:architecture}
\end{figure}

For $x_i\ne0$, define gross exposure $G_i=\norm{x_i}_1$, nominal breadth $\ell_i=|\{j:x_{ij}>0\}|$, the Herfindahl concentration index of exposure $H_i=\norm{x_i}_2^2/G_i^2$, and effective breadth $N_i^{\mathrm{eff}}=1/H_i$.  Effective and nominal breadth coincide only for an equal split.

These quantities separate scale from composition.  Gross exposure $G_i$ records how many units of slashable commitments the capital pool supports.  The pair $(H_i,N_i^{\mathrm{eff}})$ records how evenly those commitments are distributed.  In the three-position example from the introduction, equal unique stake has $G_i=W_i$ and $N_i^{\mathrm{eff}}=3$, whereas full shared collateral has $G_i=3W_i$ and the same effective breadth.  The portfolios are equally spread, but they do not create the same amount of claim.

Operator-level delegation fits the model without changing the economic object being optimized.  Let $a_i\ge0$ allocate capital across operators, with $\1^{\trans}a_i\le W_i$, and let the nonnegative matrix $Q$ map one delegated unit into slashable-position exposure:
\begin{equation}
\label{eq:operator-map}
x_i^{\trans}=a_i^{\trans}Q.
\end{equation}
The row sum of $Q$ records how much gross exposure one delegated unit creates.
\begin{proposition}[Protocol architecture induces a reuse cap]
\label{prop:architecture-map}
Assume $0\le Q_{hj}\le1$ and define
\begin{equation}
\label{eq:kappa-q}
\kappa(Q)=\max\left\{1,\max_h\sum_jQ_{hj}\right\}.
\end{equation}
Then the induced exposure satisfies $x_{ij}\le W_i$ and
$\1^{\trans}x_i\le\kappa(Q)W_i$, so $x_i\in\cX_i(\kappa(Q))$.
Unique-stake rows satisfy $\sum_jQ_{hj}\le1$; a binary shared-collateral row with $k$ ones can create gross exposure $kW_i$ from capital $W_i$.
\end{proposition}

\begin{proof}
For every position $j$,
\[
x_{ij}=\sum_h a_{ih}Q_{hj}
\le \sum_h a_{ih}
\le W_i.
\]
Moreover,
\[
\1^{\trans}x_i
=a_i^{\trans}Q\1
=\sum_h a_{ih}\sum_jQ_{hj}
\le\kappa(Q)\1^{\trans}a_i
\le\kappa(Q)W_i.
\]
These are exactly the two constraints defining $\cX_i(\kappa(Q))$.
\end{proof}


We work in exposure space because rewards and slash risk depend on $x_i$, even when several delegation vectors induce the same exposure.  The matrix $Q$ provides the bridge back to operator-level implementation.

\subsection{Correlated gross slash claims}

We distinguish a \emph{gross slash claim} from the deduction ultimately borne by a pool.  A commitment of size $x_{ij}$ to position $j$ creates a claim of $\alpha_jx_{ij}$ when that position is slashed, where $\alpha_j\in[0,1]$ is its slash fraction.  Under shared collateral, several such claims may refer to the same underlying capital.  Their sum can therefore exceed the pool's assets; a recovery, priority, or haircut rule would determine which claims are ultimately paid.  Our loss variable records claims before that recovery stage.

Let $F=(F_1,\ldots,F_m)\in\{0,1\}^m$ be the vector of slash indicators, where $F_j=1$ if position $j$ is slashed.  Write
$p=\E[F]$, $C=\Cov(F)$, and
$D_\alpha=\diag(\alpha_1,\ldots,\alpha_m)$.
Allocator $i$'s additive gross slash claim is
\begin{equation}
\label{eq:loss}
L_i(x_i,F)=x_i^{\trans}D_\alpha F,
\qquad
d=D_\alpha p,
\qquad
V=D_\alpha C D_\alpha.
\end{equation}
Thus $\E[L_i]=d^{\trans}x_i$ and
$\Var(L_i)=x_i^{\trans}Vx_i$.

Specifying pairwise correlations directly can produce a covariance matrix that no joint Bernoulli distribution can realize.  We therefore generate slash events from shared latent factors.  Each position loads on factors representing, for example, operator control, software clients, cloud providers, or market-wide stress, together with an idiosyncratic shock.  Thresholding the resulting Gaussian variables produces binary slash events with prescribed marginal probabilities and a coherent joint distribution.

Formally, for $s$ shared latent factors, choose a loading vector $\beta_j\in\R^s$ for each position, with $\norm{\beta_j}_2\le1$.  Let $Z\sim\mathcal N(0,I_s)$, let $\varepsilon_j\sim\mathcal N(0,1)$ be independent of one another and of $Z$, and set
\begin{equation}
\label{eq:factor}
Y_j=\beta_j^{\trans}Z
+\sqrt{1-\norm{\beta_j}_2^2}\,\varepsilon_j,
\qquad
F_j=\mathbf1\{Y_j\le\Phi^{-1}(p_j)\}.
\end{equation}
Here $\Phi$ denotes the standard normal cumulative distribution function and $\mathbf1\{\cdot\}$ denotes the indicator function.  Then $\Prob(F_j=1)=p_j$ and
\begin{equation}
\label{eq:binary-cov}
\Cov(F_j,F_k)
=
\Phi_2\!\left(
\Phi^{-1}(p_j),
\Phi^{-1}(p_k);
\beta_j^{\trans}\beta_k
\right)-p_jp_k,
\end{equation}
where $\Phi_2(a,b;r)$ denotes the bivariate standard normal cumulative distribution function with correlation $r$.

\begin{proposition}[Feasible factor dependence]
\label{prop:factor}
The construction in \eqref{eq:factor} generates a valid joint Bernoulli distribution, so $C=\Cov(F)\succeq0$.  For $p_j,p_k\in(0,1)$, holding the marginal probabilities fixed, the binary covariance in \eqref{eq:binary-cov} is strictly increasing in the latent correlation $\beta_j^{\trans}\beta_k$ on $(-1,1)$.
\end{proposition}

\begin{proof}
The construction defines all coordinates of $F$ on a common probability space, so it produces a valid joint Bernoulli vector with the stated marginals.  For any $z\in\R^m$,
\[
z^{\trans}Cz=\Var(z^{\trans}F)\ge0,
\]
and hence $C\succeq0$.  Let $a=\Phi^{-1}(p_j)$, $b=\Phi^{-1}(p_k)$, and $r=\beta_j^{\trans}\beta_k$.  Plackett's identity gives
\[
\frac{\partial}{\partial r}\Phi_2(a,b;r)=\phi_2(a,b;r)>0
\]
for $r\in(-1,1)$, which proves strict monotonicity in the latent correlation; continuity extends the ordering to attainable boundary cases~\citep{plackett1954}.
\end{proof}

Although the latent Gaussian variables are generated by a small number of common factors, the resulting binary slash indicators generally do not inherit the same low-rank covariance structure.  

\subsection{Rewards and private payoffs}

Position $j$ offers a fixed reward pool $R_j>0$ and has background exposure $b_j>0$.  The background term can represent passive stake, incumbents outside the modeled strategic set, or a protocol-supplied baseline.  It ensures that strategic allocators compete both with one another and with stake that they do not control.  If $X_j=\sum_hx_{hj}$ is total strategic exposure, allocator $i$ receives
\begin{equation}
\label{eq:reward}
r_{ij}(x)=R_j\frac{x_{ij}}{b_j+X_j}.
\end{equation}
The fraction $x_{ij}/(b_j+X_j)$ is the allocator's share of all exposure at that position.  Additional exposure raises its reward, but at a decreasing rate because it dilutes the reward earned by every unit already present.

Allocator $i$'s baseline payoff is
\begin{equation}
\label{eq:payoff}
u_i(x)
=
\sum_{j=1}^mR_j\frac{x_{ij}}{b_j+X_j}
-d^{\trans}x_i
-\frac{\eta_i}{2}x_i^{\trans}Vx_i,
\end{equation}
where $\eta_i\ge0$ is a reduced-form weight on variance.  The three terms are proportional reward, expected gross slash claims, and a penalty for uncertain claims.  Base-layer staking rewards are omitted because they are constant across restaking allocations; operator fees can be absorbed into net rewards or linear costs.  

\subsection{System-level quantities}

Let $X=\sum_i x_i$ be aggregate exposure.  Total reward received by strategic capital and aggregate gross claims are
\begin{equation}
\label{eq:system-quantities}
\mathsf{Reward}(x)=\sum_jR_j\frac{X_j}{b_j+X_j},
\qquad
L^{\mathrm{agg}}=X^{\trans}D_\alpha F.
\end{equation}
Consequently, $\E[L^{\mathrm{agg}}]=d^{\trans}X$ and $\Var(L^{\mathrm{agg}})=X^{\trans}VX$.  Expected gross claims add across allocators, but variance satisfies
\begin{equation}
\label{eq:aggregate-risk}
\Var(L^{\mathrm{agg}})=\sum_i x_i^{\trans}Vx_i+2\sum_{i<k}x_i^{\trans}Vx_k.
\end{equation}
The cross terms capture common-shock risk across pools.  Each allocator accounts for the variance of its own gross claims, but not for their covariance with claims generated by other allocators.  The terms $2x_i^{\trans}Vx_k$ capture this additional system-level risk.  We therefore report private utility and aggregate risk separately.

\begin{table}[tbp]
\caption{Core model objects.  Architecture determines the feasible exposure set; the reward and loss primitives determine how an exposure profile is evaluated.}
\label{tab:notation}
\centering
\small
\begin{tabularx}{\linewidth}{@{}lX@{}}
\toprule
Symbol & Meaning \\
\midrule
$W_i$ & Capital controlled by allocator $i$ \\
$x_{ij}$, $X_j$ & Allocator $i$'s exposure and aggregate strategic exposure at position $j$ \\
$\kappa_i$ & Reuse cap, limiting gross exposure to $\kappa_iW_i$ \\
$G_i$, $H_i$, $N_i^{\mathrm{eff}}$ & Gross exposure, concentration, and effective breadth \\
$F$, $p$, $C$ & Slash indicators, marginal probabilities, and their covariance \\
$D_\alpha$, $d$, $V$ & Slash fractions, expected-loss vector, and loss covariance \\
$R_j$, $b_j$ & Reward pool and nonstrategic background exposure at position $j$ \\
$\eta_i$ & Allocator $i$'s variance-penalty weight \\
\bottomrule
\end{tabularx}
\end{table}

\section{Collateral Architecture and Risk}
\label{sec:risk}

We first set rewards and strategic interaction aside to isolate the effect of collateral architecture on risk.  The analysis proceeds in three steps.  We decompose variance into gross exposure, concentration, and dependence; compare unique stake with full reuse and derive minimum and maximum variance under intermediate reuse caps; and then study how risk sensitivity changes the decision to add another position under the two endpoint architectures.

Throughout this section, positions have common slash fraction $\alpha$, marginal probability $p$, and attainable pairwise indicator correlation $\rho\in[0,1)$.  Set $\sigma^2=\alpha^2p(1-p)$.

\subsection{Gross exposure, concentration, and dependence}

\begin{theorem}[Exposure--concentration decomposition]
\label{thm:decomposition}
For every nonzero $x\ge0$,
\begin{equation}
\label{eq:decomposition}
\Var(L(x))
=
\sigma^2\left[(1-\rho)\norm{x}_2^2+\rho\norm{x}_1^2\right]
=
\sigma^2G^2\left[\rho+(1-\rho)H\right].
\end{equation}
If $x$ has $\ell$ active positions, then
\begin{equation}
\label{eq:breadth-bounds}
\sigma^2G^2\left[\rho+\frac{1-\rho}{\ell}\right]
\le
\Var(L(x))
\le
\sigma^2G^2,
\end{equation}
with the lower bound attained by equal exposure.
\end{theorem}

\begin{proof}
Under equicorrelation,
\[
C=p(1-p)\bigl[(1-\rho)I+\rho\1\1^{\trans}\bigr].
\]
Therefore
\[
\Var(L(x))
=\alpha^2x^{\trans}Cx
=\sigma^2\bigl[(1-\rho)\norm{x}_2^2+\rho(\1^{\trans}x)^2\bigr],
\]
which yields \eqref{eq:decomposition} after substituting $G=\norm{x}_1$ and $H=\norm{x}_2^2/G^2$.  If $x$ has $\ell$ active coordinates, Cauchy--Schwarz gives $G^2\le\ell\norm{x}_2^2$, with equality at an equal split.  Nonnegativity gives $\norm{x}_2^2\le G^2$.  Substitution gives \eqref{eq:breadth-bounds}.
\end{proof}

\paragraph{Reading the decomposition.}
The term $\sigma^2\rho G^2$ depends only on gross exposure and cannot be diversified away.  Breadth reduces only the idiosyncratic term $\sigma^2(1-\rho)\norm{x}_2^2$.  Equivalently,
$\Var(L)=\sigma^2G^2[\rho+(1-\rho)/N^{\mathrm{eff}}]$, so adding positions matters only insofar as exposure is actually spread across them.

For capital $W$ and $\ell$ active positions, compare equal unique stake $x^U=(W/\ell)\1_\ell$ with full shared collateral $x^S=W\1_\ell$.

\begin{corollary}[Architecture endpoints]
\label{cor:endpoints}
\begin{subequations}
\begin{align}
\E[L(x^U)]&=\alpha pW,
&\Var(L(x^U))&=\sigma^2W^2\left[\rho+\frac{1-\rho}{\ell}\right],
\label{eq:unique-risk}\\[-0.25ex]
\E[L(x^S)]&=\alpha pW\ell,
&\Var(L(x^S))&=\sigma^2W^2\ell[1+(\ell-1)\rho].
\label{eq:shared-risk}
\end{align}
\end{subequations}
Hence $\Var(L(x^S))=\ell^2\Var(L(x^U))$.  Treating slash events as independent understates either variance by $1+(\ell-1)\rho$.
\end{corollary}

\begin{proof}
For $x^U$, gross exposure is $W$ and $\norm{x^U}_2^2=W^2/\ell$.  For $x^S$, gross exposure is $\ell W$ and $\norm{x^S}_2^2=\ell W^2$.  Substituting these quantities into \eqref{eq:decomposition} gives the two formulas.  Their ratio is $\ell^2$, and setting $\rho=0$ removes the multiplicative factor $1+(\ell-1)\rho$ from either expression.
\end{proof}

The two cases use the same dependence structure.  The difference is that unique stake keeps $G=W$, whereas full reuse makes $G=\ell W$.  Under unique stake, diversification reduces the position-specific component of risk, but positive correlation leaves a residual level of variance that cannot be diversified away.  Under full reuse, by contrast, breadth also increases gross exposure, so correlation amplifies the resulting growth in risk.

\subsection{Risk bounds within the collateral-reuse envelope}

The collateral-reuse envelope spans the two architecture endpoints.  To characterize the range of risk it permits at an intermediate cap, condition on exhausting that cap, so $G=\kappa W$, and let $q=\lfloor\kappa\rfloor$ and $r=\kappa-q$.
\begin{proposition}[Risk envelope under bounded reuse]
\label{prop:risk-envelope}
Under equicorrelated slashing,
\begin{align}
\min_{x\in\cX(\kappa):\,\norm{x}_1=\kappa W}\Var(L(x))
&=\sigma^2W^2\kappa^2\left[\rho+\frac{1-\rho}{m}\right],
\label{eq:risk-envelope-min}\\
\max_{x\in\cX(\kappa):\,\norm{x}_1=\kappa W}\Var(L(x))
&=\sigma^2W^2\left[(1-\rho)(q+r^2)+\rho\kappa^2\right].
\label{eq:risk-envelope-max}
\end{align}
The minimum spreads exposure equally over all positions; the maximum fills $q$ positions to $W$ and, when $r>0$, places the remainder $rW$ on one additional position.
\end{proposition}

\begin{proof}
At fixed $G=\kappa W$, \eqref{eq:decomposition} differs across feasible portfolios only through $\norm{x}_2^2$.  Cauchy--Schwarz gives
\[
\norm{x}_2^2\ge\frac{(\1^{\trans}x)^2}{m}
=\frac{\kappa^2W^2}{m}.
\]
Equality is attained by $x_j=\kappa W/m$ for every $j$, which is feasible because $\kappa\le m$.

For the maximum, suppose two coordinates lie strictly between $0$ and $W$.  Moving a small amount from the smaller coordinate to the larger one preserves their sum and increases the sum of their squares.  Repeating this exchange leaves at most one fractional coordinate.  The maximizing vector therefore has $q=\lfloor\kappa\rfloor$ entries equal to $W$, one additional entry equal to $rW$ when $r>0$, and the rest equal to zero, so $\norm{x}_2^2=W^2(q+r^2)$.  Substituting the minimizing and maximizing squared norms into \eqref{eq:decomposition} proves the result.
\end{proof}

Thus the cap controls gross exposure, while the allocation inside the cap controls concentration.  For integer $\kappa$, worst-case variance is $\sigma^2W^2\kappa[1+(\kappa-1)\rho]$ and grows quadratically under positive correlation.

\subsection{Breadth and risk sensitivity}

Architecture also reverses the effect of risk sensitivity on breadth.  Consider symmetric positions with reward pool $R$, background exposure $B$, expected loss $d=\alpha p$, a fixed participation cost $c$ for each active position, and capital $W$ already committed to restaking.  Under unique stake, equal allocation across $\ell$ positions has value
\begin{equation}
\label{eq:unique-value}
\Pi_\ell^U
=
\frac{RW\ell}{B\ell+W}-dW
-\frac{\eta\sigma^2W^2}{2}\left[\rho+\frac{1-\rho}{\ell}\right]-c\ell.
\end{equation}
Under full reuse,
\begin{equation}
\label{eq:shared-value}
\Pi_\ell^S
=
\ell\left(\frac{RW}{B+W}-dW-c\right)
-\frac{\eta\sigma^2W^2}{2}\ell[1+(\ell-1)\rho].
\end{equation}

The comparison is conditional on capital already participating in restaking.  The choice is how many positions should carry that capital, not whether the allocator should participate at all.  A fixed per-position cost makes breadth an economically meaningful discrete choice.  Under unique stake, another position spreads a fixed exposure budget more finely.  Under full reuse, another position creates an additional full-pool claim.

\begin{proposition}[Architecture reversal]
\label{prop:reversal}
The marginal values of adding position $\ell+1$ are
\begin{align}
\Delta_\ell^U
&=
\frac{RW^2}{(B\ell+W)(B(\ell+1)+W)}
+\frac{\eta\sigma^2W^2(1-\rho)}{2\ell(\ell+1)}-c,
\label{eq:delta-u}\\
\Delta_\ell^S
&=
\frac{RW}{B+W}-dW-c
-\frac{\eta\sigma^2W^2}{2}(1+2\rho\ell).
\label{eq:delta-s}
\end{align}
Both are nonincreasing in $\ell$; $\Delta_\ell^U$ is strictly decreasing, and $\Delta_\ell^S$ is strictly decreasing when $\eta\rho>0$.  Increasing risk sensitivity $\eta$ raises $\Delta_\ell^U$ but lowers $\Delta_\ell^S$.
\end{proposition}

\begin{proof}
Conditional on selecting $\ell$ symmetric positions under unique stake, an equal split maximizes the sum of concave rewards and minimizes variance at fixed gross exposure.  Subtracting \eqref{eq:unique-value} at adjacent breadths gives \eqref{eq:delta-u}.  Under full reuse, subtracting \eqref{eq:shared-value} gives \eqref{eq:delta-s}.  The reward term in $\Delta_\ell^U$ and its diversification term are strictly decreasing in $\ell$.  The shared-collateral increment is nonincreasing and is strictly decreasing when $\eta\rho>0$.  Differentiating the two expressions with respect to $\eta$ yields the stated reversal.
\end{proof}

Thus greater risk sensitivity encourages a broader portfolio when breadth diversifies a fixed budget, but discourages breadth when every position creates another full-stake claim.

\begin{table}[tbp]
\caption{Equal exposure to $\ell$ homogeneous positions under the endpoint architectures.  The dependence structure is identical; the difference is how architecture maps capital into gross exposure.}
\label{tab:architecture-comparison}
\centering
\small
\begin{tabularx}{\linewidth}{@{}l>{\raggedright\arraybackslash}p{.22\linewidth}>{\raggedright\arraybackslash}p{.25\linewidth}X@{}}
\toprule
Quantity & Unique stake & Full shared collateral & Interpretation \\
\midrule
Exposure per active position & $W/\ell$ & $W$ & Partition versus reuse \\
Gross exposure $G$ & $W$ & $\ell W$ & Fixed budget versus collateral multiplicity \\
Expected gross claim & $\alpha pW$ & $\alpha pW\ell$ & Reuse adds expected claims \\
Variance & $\sigma^2W^2[\rho+(1-\rho)/\ell]$ & $\sigma^2W^2\ell[1+(\ell-1)\rho]$ & Floor versus quadratic growth \\
Effect of higher $\eta$ on breadth & Weakly expands & Weakly contracts & Diversification versus leverage \\
\bottomrule
\end{tabularx}
\end{table}

\section{Individual Allocation}
\label{sec:individual}

We begin with the decision faced by one allocator.  This isolates the portfolio tradeoff before introducing strategic feedback.  Hold the other allocators fixed and let $B_j>0$ denote their exposure plus background stake at position $j$.  The allocator solves
\begin{equation}
\label{eq:single-objective}
U(x)=\sum_jR_j\frac{x_j}{B_j+x_j}-d^{\trans}x-\frac{\eta}{2}x^{\trans}Vx,
\qquad x\in\cX(\kappa).
\end{equation}

\begin{theorem}[Unique individual optimum]
\label{thm:single}
If $R_j,B_j>0$ for every $j$, $V\succeq0$, and $\eta\ge0$, then $U$ is strictly concave and has a unique maximizer.  At any coordinate satisfying $0<x_j<W$, the KKT conditions imply
\begin{equation}
\label{eq:kkt}
\frac{R_jB_j}{(B_j+x_j)^2}
-d_j
-\eta(Vx)_j
=
\lambda,
\end{equation}
where $\lambda\ge0$ is the multiplier on the gross-exposure constraint $\1^{\trans}x\le\kappa W$.
\end{theorem}

\begin{proof}
The reward Hessian is diagonal with entries
\[
-\frac{2R_jB_j}{(B_j+x_j)^3}<0.
\]
Expected loss is linear, and the risk Hessian is $-\eta V\preceq0$.  Hence $U$ is strictly concave on the compact convex feasible set and has a unique maximizer.  The KKT stationarity conditions reduce to \eqref{eq:kkt} on coordinates satisfying $0<x_j<W$.
\end{proof}

Equation~\eqref{eq:kkt} balances marginal reward against three marginal costs: the expected slash cost, the variance penalty, and the shadow price of scarce gross-exposure capacity.  The variance term satisfies
\[
(Vx)_j
=
V_{jj}x_j+\sum_{k\ne j}V_{jk}x_k,
\]
so it accounts for both position $j$'s own variance and its covariance with the existing portfolio; the corresponding marginal penalty is $\eta(Vx)_j$. Thus a position with low individual slash risk may still be unattractive if it shares operators, software, or infrastructure with positions already in the portfolio, because it increases losses precisely in states where the portfolio is already losing.

The multiplier $\lambda$ is the marginal value of one additional unit of gross-exposure capacity.  To express the value of relaxing the reuse cap itself, define
\[
v(\kappa)=\max_{x\in\cX(\kappa)}U(x).
\]

\begin{proposition}[Value of reuse]
\label{prop:reuse-value}
The value function $v$ is nondecreasing and concave on $[1,m]$.  At a differentiability point,
\begin{equation}
\label{eq:reuse-value}
v'(\kappa)=\lambda^\star W,
\end{equation}
where $\lambda^\star$ is the optimal multiplier on the aggregate exposure constraint.
\end{proposition}

\begin{proof}
The feasible sets are nested in $\kappa$, so $v$ is nondecreasing.  If $x^1$ and $x^2$ are optimal at $\kappa_1$ and $\kappa_2$, respectively, then $tx^1+(1-t)x^2$ is feasible at $t\kappa_1+(1-t)\kappa_2$.  Concavity of $U$ therefore implies concavity of $v$.
The dependence on $\kappa$ enters through the aggregate exposure constraint
\[
\1^{\trans}x\le\kappa W.
\]
Its Lagrangian is
\[
\mathcal L(x,\lambda;\kappa)
=
U(x)+\lambda\bigl(\kappa W-\1^{\trans}x\bigr).
\]
At a differentiability point, the envelope theorem therefore gives
\[
v'(\kappa)
=
\frac{\partial\mathcal L}{\partial\kappa}
=
\lambda^\star W,
\]
which proves \eqref{eq:reuse-value}.
\end{proof}

This shadow value is private: it does not account for the reward-displacement or cross-pool common-shock externalities introduced below.

\begin{proposition}[Regret from ignoring positive dependence]
\label{prop:regret}
Let $V_0$ be the covariance model used by the allocator and let
$V=V_0+\Delta$ be the true covariance matrix.  Suppose
$x^{\trans}\Delta x\ge0$ for every feasible exposure vector $x$; this holds, for example, when every entry of $\Delta$ is nonnegative.  Let $x^0$ maximize the individual objective using $V_0$, and let $x^\star$ maximize the true objective using $V$.  Then
\begin{equation}
\label{eq:regret-bound}
0\le U_V(x^\star)-U_V(x^0)
\le \frac{\eta}{2}(x^0)^{\trans}\Delta x^0.
\end{equation}
\end{proposition}

\begin{proof}
Write $U_V(x)=U_{V_0}(x)-(\eta/2)x^{\trans}\Delta x$.  Optimality of $x^0$ under $V_0$ gives $U_{V_0}(x^\star)-U_{V_0}(x^0)\le0$.  Therefore
\[
U_V(x^\star)-U_V(x^0)
\le\frac{\eta}{2}\left[(x^0)^{\trans}\Delta x^0-(x^\star)^{\trans}\Delta x^\star\right]
\le\frac{\eta}{2}(x^0)^{\trans}\Delta x^0.
\]
The lower bound follows from optimality of $x^\star$ under the true objective.
\end{proof}

The upper bound is the omitted covariance penalty evaluated at the misspecified portfolio.  It motivates comparing allocations chosen under the true covariance structure with those chosen under independence or an estimated covariance model.  A more elaborate dependence estimate is useful only if the reduction in misspecification outweighs the estimation noise it introduces.

The mean--variance specification is not essential.  Replacing the combined expected-claim and variance terms with any differentiable convex private risk cost preserves the reward-based monotonicity in what follows.  For example, the exact entropic risk functional is
\begin{equation}
\label{eq:entropic-risk}
\mathcal R_\theta(x)
=
\frac{1}{\theta}
\log\E\!\left[e^{\theta x^{\trans}D_\alpha F}\right],
\end{equation}
where $\theta>0$ is a risk-sensitivity parameter.  This functional is convex, and its small-$\theta$ expansion is
\[
\mathcal R_\theta(x)
=
\E[L(x)]
+\frac{\theta}{2}\Var(L(x))
+O(\theta^2).
\]

\section{Strategic Competition and Equilibrium}
\label{sec:equilibrium}

We now return to the $n$-allocator game.  In a game there is generally no single objective whose gradient governs every participant.  The game's \emph{pseudo-gradient} stacks each allocator's payoff gradient with respect to its own decision.  We work with its negative, denoted by $\cF$:
\begin{equation}
\label{eq:pseudogradient}
g(x)=(\nabla_{x_1}u_1(x),\ldots,\nabla_{x_n}u_n(x)),
\qquad \cF(x)=-g(x).
\end{equation}
A profile is an equilibrium precisely when no feasible direction of unilateral adjustment has positive first-order value.  Write $\cX=\prod_i\cX_i(\kappa_i)$.  Equivalently, an equilibrium solves the variational inequality
\begin{equation}
\label{eq:vi}
\langle\cF(x^\star),x-x^\star\rangle\ge0
\qquad\text{for every }x\in\cX.
\end{equation}
Let $\overline W=\sum_iW_i$ and
\begin{equation}
\label{eq:mu}
\mu_R=\min_j\frac{R_jb_j}{(b_j+\overline W)^3}>0.
\end{equation}

\begin{theorem}[Unique equilibrium and certified computation]
\label{thm:equilibrium}
Suppose $R_j,b_j>0$, $V\succeq0$, and $\eta_i\ge0$.  Then $\cF$ is strongly monotone with modulus at least $\mu_R$:
\begin{equation}
\label{eq:strong-monotonicity}
\langle\cF(x)-\cF(y),x-y\rangle
\ge\mu_R\norm{x-y}_2^2.
\end{equation}
Consequently, the game has a unique pure Nash equilibrium.  If $L_{\cF}$ is a Lipschitz constant of $\cF$ on $\cX$, then
\begin{equation}
\label{eq:pg}
x^{t+1}=\Pi_{\cX}(x^t-\tau\cF(x^t)),
\qquad
0<\tau<\frac{2\mu_R}{L_{\cF}^2},
\end{equation}
converges globally and linearly to that equilibrium.
\end{theorem}

The proof has four steps.  First, positive background exposure gives the proportional-reward game a uniform amount of curvature, even though the game need not admit a potential.  Second, private risk costs preserve that curvature and make $\cF$ strongly monotone.  Third, strong monotonicity rules out two distinct equilibria.  Finally, strong monotonicity and Lipschitz continuity make a sufficiently small projected step a contraction.

\begin{proof}
\emph{Step 1: one-position reward curvature.}
Fix one position, suppress its index, and write
\[
q=(q_1,\ldots,q_n),
\qquad
T=b+\1^{\trans}q.
\]
Player $i$'s own marginal reward is
\[
g_i^R(q)=R\frac{T-q_i}{T^2}.
\]
Its Jacobian has entries
\[
\frac{\partial g_i^R}{\partial q_i}
=-\frac{2R(T-q_i)}{T^3},
\qquad
\frac{\partial g_i^R}{\partial q_k}
=\frac{R(2q_i-T)}{T^3}\quad(k\ne i).
\]
Let $S^R=(J^R+(J^R)^{\trans})/2$ be the symmetric part.  For an arbitrary direction $z$, set $a=\1^{\trans}z$ and $c=q^{\trans}z$.  Direct expansion gives
\begin{equation}
\label{eq:reward-quadratic}
z^{\trans}S^R(q)z
=\frac{R}{T^3}
\left[-T\bigl(a^2+\norm{z}_2^2\bigr)+2ac\right].
\end{equation}
Because $q\ge0$, $\norm{q}_2\le\norm{q}_1=T-b$, and hence $|c|\le(T-b)\norm{z}_2$.  It follows that
\begin{align*}
T\bigl(a^2+\norm{z}_2^2\bigr)-2ac
&\ge T\bigl(a^2+\norm{z}_2^2\bigr)
-2(T-b)|a|\norm{z}_2\\
&=b\bigl(a^2+\norm{z}_2^2\bigr)
+(T-b)\bigl(|a|-\norm{z}_2\bigr)^2\\
&\ge b\norm{z}_2^2.
\end{align*}
Combining this inequality with \eqref{eq:reward-quadratic} yields
\begin{equation}
\label{eq:symmetric-jacobian-bound}
z^{\trans}S^R(q)z
\le-\frac{Rb}{T^3}\norm{z}_2^2.
\end{equation}

\medskip\noindent
\emph{Step 2: aggregation and private risk.}
At any feasible profile, total strategic exposure to one position is at most $\overline W=\sum_iW_i$, so $T\le b+\overline W$.  Applying \eqref{eq:symmetric-jacobian-bound} position by position bounds the symmetric Jacobian of the stacked reward pseudo-gradient above by $-\mu_RI$.  Expected loss is linear.  The private variance term contributes the block $-\eta_iV\preceq0$ to player $i$'s own Jacobian, so it can only strengthen the inequality.

For arbitrary feasible $x,y$, let $z=x-y$.  Convexity of the feasible set ensures that $y+tz$ is feasible for every $t\in[0,1]$.  Integrating the Jacobian along this segment gives
\[
\langle g(x)-g(y),z\rangle
=\int_0^1z^{\trans}J_g(y+tz)z\,dt
\le-\mu_R\norm{z}_2^2.
\]
Thus $\cF=-g$ is strongly monotone with modulus at least $\mu_R$.

\medskip\noindent
\emph{Step 3: existence and uniqueness.}
Existence follows from continuity, compact convex strategy sets, and concavity in each allocator's own action.  Suppose $x^\star$ and $y^\star$ both solve \eqref{eq:vi}.  Using each solution as the comparison point for the other and adding gives
\[
\langle\cF(x^\star)-\cF(y^\star),x^\star-y^\star\rangle\le0.
\]
Strong monotonicity makes the left side at least $\mu_R\norm{x^\star-y^\star}_2^2$, so the two solutions coincide.

\medskip\noindent
\emph{Step 4: projected-method convergence.}
Every solution of \eqref{eq:vi} is a fixed point of $T_\tau(x)=\Pi_{\cX}(x-\tau\cF(x))$.  Projection is nonexpansive, and hence
\begin{align}
\norm{T_\tau(x)-T_\tau(y)}_2^2
&\le\norm{x-y-\tau(\cF(x)-\cF(y))}_2^2\notag\\
&\le\bigl(1-2\tau\mu_R+\tau^2L_{\cF}^2\bigr)
\norm{x-y}_2^2.
\label{eq:contraction-main}
\end{align}
The step-size condition in \eqref{eq:pg} makes the multiplier smaller than one, proving global linear convergence to the unique equilibrium.
\end{proof}

Economically, $b_j>0$ means that background stake competes for the reward pool even when strategic exposure is zero.  Because the reward pool is fixed, an allocator's reward rises with its own exposure but at a decreasing rate: each additional unit captures less new reward than the preceding one.  Positive background exposure makes this diminishing-return effect uniform over the feasible set, providing the curvature behind equilibrium uniqueness.  When $b_j=0$, by contrast, an arbitrarily small first deposit at an empty position captures the entire reward pool; the payoff rule is singular at zero, and this uniform curvature disappears.

\subsection{Robustness and extensions}

Strong monotonicity also makes the equilibrium robust to errors in the estimated marginal incentives.  If rewards, background exposure, or slashing risk are slightly misspecified, the resulting equilibrium cannot move arbitrarily far.
\begin{proposition}[Equilibrium sensitivity]
\label{prop:sensitivity}
Let $x^\star$ solve $\operatorname{VI}(\cX,\cF)$ and let $\widetilde x^\star$ solve $\operatorname{VI}(\cX,\widetilde{\cF})$.  Suppose $\cF$ is $\mu$-strongly monotone and
\begin{equation}
\label{eq:operator-error}
\sup_{x\in\cX}
\norm{\cF(x)-\widetilde{\cF}(x)}_2
\le\varepsilon.
\end{equation}
Then
\begin{equation}
\label{eq:sensitivity}
\norm{x^\star-\widetilde x^\star}_2
\le\frac{\varepsilon}{\mu}.
\end{equation}
\end{proposition}

\begin{proof}
Let $d=x^\star-\widetilde x^\star$.  The two variational inequalities give
$\langle\cF(x^\star),d\rangle\le0$ and
$\langle\widetilde{\cF}(\widetilde x^\star),d\rangle\ge0$.
Strong monotonicity, followed by these two inequalities, therefore gives
\begin{align*}
\mu\norm{d}_2^2
&\le
\langle\cF(x^\star)-\cF(\widetilde x^\star),d\rangle\\
&\le
\langle
\widetilde{\cF}(\widetilde x^\star)
-\cF(\widetilde x^\star),d
\rangle\\
&\le
\varepsilon\norm{d}_2,
\end{align*}
where the last inequality follows from Cauchy--Schwarz and the assumed uniform perturbation bound.  Dividing by $\mu\norm{d}_2$ when $d\ne0$ proves the result; the case $d=0$ is immediate.
\end{proof}

Thus $\varepsilon$ measures the largest error in the marginal-incentive vector, while $1/\mu$ converts that error into a worst-case displacement of the equilibrium.  A stronger restoring force (larger $\mu$) makes the allocation less sensitive to the same modeling error.

\begin{corollary}[Convex private risk costs]
\label{cor:convex-risk}
Replace the quadratic term in \eqref{eq:payoff} by any differentiable convex cost $\mathcal R_i(x_i)$.  The game still has a unique pure equilibrium.  If each $\nabla\mathcal R_i$ is Lipschitz, projected pseudo-gradient converges linearly for a sufficiently small step.  If $\mathcal R_i$ is $\gamma_i$-strongly convex, the monotonicity modulus is at least $\mu_R+\min_i\gamma_i$.
\end{corollary}

\begin{proof}
The gradient of a differentiable convex function is monotone.  Adding the blocks $\nabla\mathcal R_i$ to $\cF$ therefore preserves the reward-based modulus $\mu_R$; strong convexity adds $\min_i\gamma_i$.  Lipschitz gradients on the compact feasible set yield a finite Lipschitz constant, so the projection argument in \Cref{thm:equilibrium} applies.
\end{proof}

\begin{corollary}[Equilibrium under fixed operator maps]
\label{cor:operator-map}
Suppose allocator $i$ chooses operator delegation $a_i$ from a nonempty compact convex set $\mathcal A_i$, and induced exposure is $x_i^{\trans}=a_i^{\trans}Q$ for a fixed nonnegative matrix $Q$.  Assume $Q^{\trans}\mathcal A_i\subseteq\cX_i(\kappa_i)$ for each $i$.  Equilibrium exposure in the induced sets $Q^{\trans}\mathcal A_i$ is unique.  Operator delegations need not be unique because the map $a_i\mapsto Q^{\trans}a_i$ may be noninjective on $\mathcal A_i$.
\end{corollary}

\begin{proof}
The induced exposure sets $Q^{\trans}\mathcal A_i$ are compact and convex.  Applying \Cref{thm:equilibrium} to those sets gives a unique exposure profile.  If two distinct delegations in $\mathcal A_i$ differ by an element of $\ker(Q^{\trans})$, they induce the same exposure and payoff.
\end{proof}

\subsection{Symmetric benchmark}

In a fully symmetric market the equilibrium reduces to a scalar equation.  If all allocators and positions share $(W,\kappa,\eta,R,b,d)$ and $V=\sigma^2[(1-\rho)I+\rho\1\1^{\trans}]$, let $\bar q=\kappa W/m$ and
\begin{equation}
\label{eq:psi}
\psi(q)=R\frac{b+(n-1)q}{(b+nq)^2}-d-\eta\sigma^2[1+(m-1)\rho]q.
\end{equation}
\begin{proposition}[Symmetric equilibrium]
\label{prop:symmetric-equilibrium}
Uniqueness and symmetry imply a $q^\star$ such that $x_{ij}^\star=q^\star$ for all $i,j$.  The function $\psi$ in \eqref{eq:psi} is strictly decreasing: $q^\star=0$ if $R/b\le d$, $q^\star=\bar q$ if $\psi(\bar q)\ge0$, and otherwise $q^\star$ is the unique root of $\psi(q)=0$ in $(0,\bar q)$.  At an interior solution with $m\ge2$ and $\eta>0$, $q^\star$ decreases strictly in $\rho$.
\end{proposition}

\begin{proof}
The game is invariant under permutations of allocators and positions, so uniqueness forces every coordinate to equal some $q$.  Substitution into the marginal payoff gives \eqref{eq:psi}.  Its derivative is
\[
-\frac{R[(n+1)b+n(n-1)q]}{(b+nq)^3}
-\eta\sigma^2[1+(m-1)\rho]<0.
\]
The boundary and interior cases follow from the KKT conditions.  At an interior root, implicit differentiation gives $dq^\star/d\rho<0$.
\end{proof}

The cap matters only when the unconstrained root exceeds $\bar q$.  Away from that boundary, stronger correlation lowers exposure per position even when the set of available positions is unchanged.  The formula also supplies a transparent unit test for equilibrium code.

The same scalar condition gives additional comparative statics.  At a positive interior equilibrium, exposure increases with the reward pool $R$ and decreases with expected loss $d$, risk sensitivity $\eta$, slash variance $\sigma^2$, and---when $m\ge2$---correlation $\rho$.  Relaxing $\kappa$ has no local effect when the cap is slack; while the cap binds, $q^\star=\kappa W/m$ and exposure rises mechanically at rate $W/m$.  Thus a protocol can distinguish two regimes: a scarcity regime in which architecture directly limits commitments, and an interior regime in which rewards and risk determine exposure even though more reuse is technologically available.

\section{Planner Benchmarks and Protocol Design}
\label{sec:planner}

We use two planner benchmarks for two different externalities.  The first maximizes the sum of private utilities.  It internalizes reward displaced between strategic allocators but retains each allocator's private variance cost.  The second evaluates risk at the system level and therefore also internalizes covariance between different pools' gross claims.  Keeping the benchmarks separate distinguishes reward displacement from common-shock risk.

\subsection{Reward displacement}

Consider first a planner that maximizes the sum of the strategic allocators' utilities:
\begin{equation}
\label{eq:planner}
\begin{aligned}
\mathcal W(x)
&=
\sum_i u_i(x) \\
&=
\sum_j R_j\frac{X_j}{b_j+X_j}
-d^{\trans}X
-\sum_i\frac{\eta_i}{2}x_i^{\trans}Vx_i,
\qquad X=\sum_i x_i.
\end{aligned}
\end{equation}
The planner internalizes the effect of each allocator's exposure on the rewards received by others.  Reward captured at the expense of another strategic allocator is therefore treated as a transfer, not an increase in aggregate reward.  This benchmark retains the individual variance penalties from private utilities.  A planner concerned with aggregate system risk could instead penalize $X^{\trans}VX$, which additionally captures covariance in gross claims across allocators.

\begin{proposition}[Reward-capture externality]
\label{prop:wedge}
The difference between allocator $i$'s private marginal reward and the planner's marginal reward at position $j$ is
\begin{equation}
\label{eq:wedge}
\frac{R_j(X_j-x_{ij})}{(b_j+X_j)^2}\ge0.
\end{equation}
It is zero exactly when no other strategic allocator is exposed to that position.
\end{proposition}

\begin{proof}
Differentiate allocator $i$'s reward and total strategic reward.  Their marginal difference is
\[
\frac{R_j(b_j+X_j-x_{ij})}{(b_j+X_j)^2}
-\frac{R_jb_j}{(b_j+X_j)^2}
=\frac{R_j(X_j-x_{ij})}{(b_j+X_j)^2}.
\]
\end{proof}

A unique equilibrium can therefore still involve excessive gross exposure: each allocator values reward captured from competitors, while the planner values only newly created strategic reward.  In a single position, an allocator's extra deposit has two effects.  It increases the fraction of the reward pool captured by strategic capital, and it transfers reward share away from rival strategic allocators.  The planner values the first effect but not the second.

\paragraph{Symmetric overexposure.}
The scalar benchmark makes this comparison exact.  Under the assumptions of \Cref{prop:symmetric-equilibrium}, let $q^{\mathrm P}$ be the planner's common exposure per allocator and position.  Its interior marginal condition is
\begin{equation}
\label{eq:planner-psi}
\psi^{\mathrm P}(q)
=
\frac{Rb}{(b+nq)^2}-d
-\eta\sigma^2[1+(m-1)\rho]q=0.
\end{equation}
\begin{corollary}[Symmetric overexposure]
\label{cor:overexposure}
In the symmetric model, $q^\star\ge q^{\mathrm P}$.  If $n>1$ and the planner chooses a positive interior allocation, then $q^\star>q^{\mathrm P}$.  Equality may occur at a boundary, including when both choices bind at the same upper cap.
\end{corollary}

\begin{proof}
The two marginal functions differ by
\begin{equation}
\label{eq:symmetric-wedge}
\psi(q)-\psi^{\mathrm P}(q)
=
\frac{R(n-1)q}{(b+nq)^2}\ge0.
\end{equation}
Both marginal functions are strictly decreasing, so the private zero lies weakly to the right of the planner zero.  The inequality is strict for $n>1$ at a positive interior planner choice.  Projecting both unconstrained choices onto the common interval $[0,\bar q]$ preserves the weak ordering; equality can also occur when both choices bind at the upper cap.
\end{proof}

The result isolates competition from scale: even with total capital and risk preferences fixed, splitting capital among more strategic allocators can raise aggregate exposure because each allocator attempts to capture part of a fixed pool.

\subsection{Pricing reward displacement and common-shock risk}
The planner in \eqref{eq:planner} isolates reward competition by summing private variance costs.  A protocol concerned with joint solvency may instead use
\begin{equation}
\label{eq:systemic-planner}
\mathcal W^{\mathrm{sys}}(x)
=
\sum_jR_j\frac{X_j}{b_j+X_j}
-d^{\trans}X-\frac{\eta}{2}X^{\trans}VX,
\end{equation}
where allocators share a common risk parameter $\eta$.  This objective additionally internalizes covariance between different pools' gross claims.  Let $Y_i=X-x_i$ denote competitors' exposure.
\begin{proposition}[Architecture-aware Pigouvian benchmark]
\label{prop:pigouvian}
Define the profile-dependent charge
\begin{equation}
\label{eq:pigouvian}
T_i(x)
=
\sum_jR_jY_{ij}
\left(\frac{1}{b_j+Y_{ij}}-\frac{1}{b_j+X_j}\right)
+\eta x_i^{\trans}VY_i.
\end{equation}
The game with payoff $u_i(x)-T_i(x)$ is an exact potential game with potential $\mathcal W^{\mathrm{sys}}$.  Hence every equilibrium maximizes \eqref{eq:systemic-planner}, and aggregate exposure is unique.
\end{proposition}

\begin{proof}
Fix allocator $i$ and hold $x_{-i}$ fixed, so $Y_i=X-x_i$ is constant with respect to $i$'s deviation.  For each position,
\begin{align*}
&R_j\frac{x_{ij}}{b_j+X_j}
-R_jY_{ij}\left(\frac{1}{b_j+Y_{ij}}-\frac{1}{b_j+X_j}\right)\\
&\qquad=
R_j\frac{X_j}{b_j+X_j}
-R_j\frac{Y_{ij}}{b_j+Y_{ij}}.
\end{align*}
The second term on the right depends only on $x_{-i}$.  Symmetry of $V$ similarly gives
\[
-\frac{\eta}{2}x_i^{\trans}Vx_i
-\eta x_i^{\trans}VY_i
=-\frac{\eta}{2}X^{\trans}VX
+\frac{\eta}{2}Y_i^{\trans}VY_i,
\]
and expected gross claims decompose linearly.  Therefore there exists a function $K_i$ of competitors' actions alone such that
\[
u_i(x)-T_i(x)=\mathcal W^{\mathrm{sys}}(x)+K_i(x_{-i}).
\]
Every unilateral deviation changes adjusted utility by exactly the change in $\mathcal W^{\mathrm{sys}}$, establishing exact potentiality.

The potential is concave in the full profile because it is a concave function of the linear aggregate $X$.  More specifically,
\[
\frac{d^2}{dX_j^2}
\left(R_j\frac{X_j}{b_j+X_j}\right)
=-\frac{2R_jb_j}{(b_j+X_j)^3}<0,
\]
while expected loss is linear and $-\eta X^{\trans}VX/2$ is concave.  The first-order conditions for a Nash equilibrium of the potential game are therefore the blockwise first-order conditions for a global maximizer, and conversely every maximizer is an equilibrium.  Strict concavity in $X$ implies that all maximizing profiles share the same aggregate exposure vector, although their decomposition across allocators need not be unique.
\end{proof}

The first term prices reward share displaced from competitors; the second prices covariance between allocator $i$'s gross claims and everyone else's.  Equation~\eqref{eq:pigouvian} identifies the marginal correction needed to align private incentives with the systemic planner, rather than requiring a charge of exactly this form.  The same incentives could instead be implemented through reward haircuts, insurance premia, or architecture-sensitive capital requirements.

\section{Computational Evaluation}
\label{sec:experiments}

The experiments ask three questions suggested directly by the theory.  First, when a protocol relaxes the reuse cap, do rewards and tail risk grow at the same rate?  Second, when a fixed amount of capital is divided among more competing allocators, does equilibrium exposure move away from the planner benchmark?  Third, when does ignoring correlation across positions materially change the chosen portfolio and its true risk?

The environments are controlled synthetic stress tests designed to isolate these mechanisms.  Slash events follow the factor model in \eqref{eq:factor}; the factors capture market-wide shocks and dependencies arising from shared operators, software clients, or cloud providers.  Each central experiment changes one economic margin while holding the rest of the environment fixed: the reuse frontier changes permitted gross exposure, the competition experiment divides fixed total capital among more independently optimizing allocators, and the misspecification experiment changes only the covariance information used to choose the portfolio.  Within a fixed environment, alternative allocations are evaluated on exactly the same slash scenarios, so their differences are not driven by different simulated shocks.

\begin{table}[H]
\caption{Design of the three central experiments.  Tail outcomes are evaluated on common simulated slash scenarios within each environment.}
\label{tab:main-experiment-design}
\centering
\small
\begin{tabularx}{\linewidth}{@{}p{.18\linewidth}p{.28\linewidth}X@{}}
\toprule
Experiment & Main variation & Fixed design and replication \\
\midrule
Reuse frontier
& $\kappa=1,\ldots,12$; three factor regimes
& $n=4$, $m=12$, five seeds, $80{,}000$ evaluation draws; 180 equilibria \\
\addlinespace
Competition
& $n\in\{1,2,4,8,12\}$; three background scales
& $m=8$, total capital $4$, $\kappa=2$, five seeds; 75 equilibria \\
\addlinespace
Misspecification
& True covariance versus diagonal and rank-three estimates
& $m=16$, $\kappa=4$, ten seeds, $120{,}000$ evaluation draws; $250$--$5{,}000$ estimation samples \\
\bottomrule
\end{tabularx}
\end{table}

\subsection{Computational procedure and reproducibility}

For each parameter setting, we compute an equilibrium using simultaneous adaptive projected pseudo-gradient updates: all allocators move in the direction of their current marginal payoffs, after which their choices are projected back onto their feasible reuse sets.  Every reported profile is checked using unilateral exploitability---the largest utility gain available from a deviation by one allocator---and a fixed-step variational-inequality (VI) residual, which measures violation of the equilibrium first-order conditions.  Both quantities vanish at an exact equilibrium.  Cyclic best response provides an independent solver check.

The accompanying artifact\footnote{\url{https://github.com/Kkenize/liquid-restaking.git}} fixes all random seeds, regenerates the processed outputs and every figure and table, and passes nine unit tests plus six higher-sample validation checks.  Across the 180 architecture profiles, the largest exploitability is $5.96\times10^{-14}$ and the largest VI residual is $4.34\times10^{-9}$.  Across the 75 competition profiles, the corresponding maxima are $3.82\times10^{-17}$ and $8.92\times10^{-9}$, so the reported profiles are numerically indistinguishable from equilibrium at the displayed precision.  High-sample Monte Carlo checks give maximum absolute errors of $4.36\times10^{-4}$ for factor covariance, $2.36\times10^{-4}$ for the mean gross claim, and $4.17\times10^{-5}$ for its variance.

Our main measure of tail severity is $99\%$ conditional value at risk (CVaR, also called expected shortfall): after sorting the simulated aggregate gross claims, we average those in the worst one percent of outcomes.  Higher CVaR therefore means that extreme claims are more severe on average, not that a slash is more likely.  Because this is a gross-claim measure before recovery or haircuts, it can exceed underlying capital when collateral is reused.

\begin{figure}[H]
\centering
\includegraphics[width=.98\linewidth]{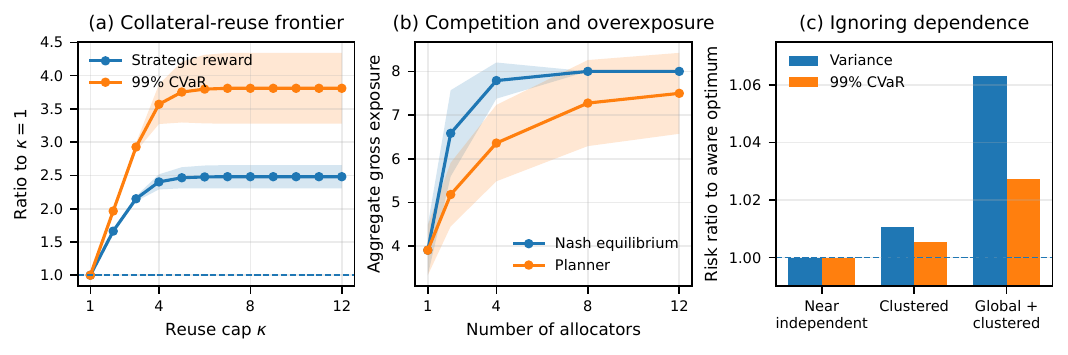}
\caption{Central computational results.  Bands in panels (a) and (b) show mean $\pm 1.96$ standard errors across five fixed heterogeneous environments.  (a) Relative to the $\kappa=1$ outcome, reuse raises $99\%$ CVaR faster than strategic reward in the global-plus-clustered regime.  (b) With total strategic capital fixed, competition produces more gross exposure than the planner.  (c) Portfolios chosen while ignoring off-diagonal covariance are evaluated under the true dependence model; a risk ratio above one means that ignoring covariance raises risk relative to the covariance-aware allocation.}
\label{fig:empirical-summary}
\end{figure}

\subsection{The collateral-reuse frontier}

With four allocators and twelve positions, we vary $\kappa$ from $1$ to $12$ under near-independent, clustered, and global-plus-clustered dependence.  This traces equilibrium choices across the collateral-reuse envelope; unlike \Cref{prop:risk-envelope}, it does not condition on exhausting the cap.  Within each dependence regime, reward and CVaR are reported as multiples of their own $\kappa=1$ values.  The multiples therefore describe how the two quantities grow with reuse.

At the largest cap, reward rises by $3.00\times$, $2.85\times$, and $2.47\times$, respectively, while aggregate CVaR rises by $4.60\times$, $4.45\times$, and $3.77\times$.  The equilibrium does not use all permitted reuse: under the strongest common factor, gross exposure reaches $3.96$ times capital, well below the permitted limit of twelve times capital.  The flattening of the curves therefore means that further relaxation of the cap no longer changes the chosen portfolio.  Effective breadth---the equally weighted position count corresponding to the portfolio's concentration---grows much less than gross exposure, so reuse primarily enlarges claims rather than proportionately diversifying them.

\subsection{Competition versus the planner}

We hold total strategic capital at four units, set $\kappa=2$, and vary the number of allocators.  Thus at most eight units of aggregate gross exposure are feasible, regardless of how many allocators divide the capital.  Here the planner is the reward-displacement benchmark: it maximizes the sum of private utilities and therefore does not value reward that is merely transferred between strategic allocators, while retaining their individual risk penalties.

At the intermediate background scale, equilibrium and planner exposure both equal $3.907$ with one allocator, when no reward can be captured from a strategic rival.  With four allocators, equilibrium exposure is $7.789$ versus $6.358$ for the planner, and aggregate $99\%$ CVaR is $2.029$ versus $1.604$.  With twelve allocators, exposure is $8.000$ versus $7.497$, and CVaR is $2.207$ versus $1.890$.  Competition therefore creates more gross exposure and more severe tail outcomes even though total underlying capital is unchanged.  The aggregate-utility gap is modest, but the difference in the loss tail is economically visible.

\subsection{Dependence misspecification}

A single allocator chooses among sixteen positions with $\kappa=4$.  We compare the covariance-aware portfolio with one chosen after setting all off-diagonal covariance terms to zero, so that the allocation rule treats positions as uncorrelated.  Both portfolios are then evaluated under the true covariance and on the same simulated slash scenarios, isolating the effect of the dependence information used when choosing the portfolio.

Under clustered factors, ignoring dependence raises true variance by $1.05\%$ and $99\%$ CVaR by $0.54\%$.  Under a strong global factor, the increases are $6.32\%$ and $2.72\%$.  Modeling dependence is therefore most valuable when common shocks are important, but its benefit also depends on estimation quality.  With 250 observations, estimation noise can cause a fitted rank-three covariance model to produce a worse allocation than the diagonal benchmark; with 5,000 observations, the same low-rank approach substantially improves allocation quality.

\subsection{Robustness}

The appendix examines whether the conclusions depend on how the common factors are grouped, which equilibrium solver is used, or which risk objective is optimized.  Cyclic, reverse-order, and randomized best responses converge in all 24 tested environments, typically within six to thirteen complete passes through the allocators.  Adaptive projected updates also converge throughout and attain the tightest VI residuals; conservative fixed steps and the $2/\sqrt{t}$ step-size schedule are less reliable under stress.  These adaptive and best-response convergence results are computational findings; the global convergence theorem applies to the certified fixed-step method under its stated step-size condition.

Mean--variance allocations and allocations under the entropic objective, which places increasing weight on large claims, nearly coincide at low risk sensitivity.  Stronger tail penalties reduce gross exposure and can make nonparticipation optimal.

Taken together, these controlled synthetic experiments show that the theoretical mechanisms persist across heterogeneous environments.  They provide mechanism evidence and benchmarks for richer protocol-specific calibrations, rather than estimates of current protocol parameters.

\section{Implications, Scope, and Limitations}
\label{sec:discussion}

\paragraph{Architecture is a first-order risk control.}
Correlation alone does not determine whether breadth is safe.  Under unique stake, breadth reallocates fixed exposure and diversifies idiosyncratic risk.  Under reuse, breadth may create additional full-stake claims.  Protocol dashboards should therefore report underlying capital $W$, gross exposure $G$, the reuse ratio $G/W$, and effective breadth $N^{\mathrm{eff}}$ separately.  A count of services or operators can make a concentrated or highly reused portfolio appear diversified.

\paragraph{Unique stake is not independence.}
Partitioning collateral prevents two positions from claiming the same unit of capital, but does not remove common operational dependencies.  Conversely, low correlation does not make unlimited reuse safe: independent claims still accumulate expected loss and can exhaust the pool.  Architecture determines how much claim is created from a unit of capital; the dependence model determines how likely those claims are to arrive together.  A sound risk assessment must report both dimensions.

\paragraph{Security procurement versus reward transfer.}
The planner comparison clarifies what proportional rewards purchase.  The system-level marginal reward from more exposure at position $j$ is $R_jb_j/(b_j+X_j)^2$.  Allocator $i$ additionally values the share transferred from competitors, exactly the wedge in \eqref{eq:wedge}.  When background exposure is small and several strategic pools compete, a large part of the private incentive to deposit can therefore be redistributive rather than an increase in total reward captured by strategic capital.  More equilibrium exposure need not represent a commensurate increase in economically useful security, especially when the same collateral is reused.

\begin{table}[tbp]
\caption{Protocol levers target different margins of the model.}
\label{tab:design-levers}
\centering
\small
\begin{tabularx}{\linewidth}{@{}p{.22\linewidth}p{.28\linewidth}X@{}}
\toprule
Lever & Model object & Primary effect \\
\midrule
Reuse cap & $\kappa$, feasible $G$ & Limits collateral multiplicity and gross claims \\
Reward design & $R_j,b_j$ & Changes entry, reward displacement, and overexposure \\
Risk charge & $(Vx_i)_j$ or $(VX)_j$ & Prices marginal common-factor exposure \\
Recovery rule & Gross-to-realized loss map & Allocates shortfall after capital is exhausted \\
\bottomrule
\end{tabularx}
\end{table}

\paragraph{Operationalizing the framework.}
A protocol can apply the model without a long history of realized slashes.  First, reconstruct $Q$ from operator-set magnitudes or service portfolios and report both capital and gross exposure.  Second, group positions by observable common dependencies---operator control, software client, cloud provider, bridge, or governance module---and translate those overlaps into conservative factor-loading scenarios.  Third, combine those scenarios with reward schedules, background exposure, and slash fractions to solve the private equilibrium and planner counterfactual.  A compact dashboard can then report $G/W$, effective breadth, expected gross claims, aggregate variance, tail claims, and the reward-capture wedge.  Scenario ranges and shrinkage are preferable to a falsely precise pairwise covariance estimate when slash events are rare.

\paragraph{Gross claims and the incidence of shortfall.}
Under reuse, simultaneous nominal claims can exceed available capital.  A waterfall, priority rule, or proportional haircut transforms gross claims nonlinearly and determines which service, insurer, or token holder absorbs the shortfall.  Such a rule may cap deductions from the liquid-restaking pool, but it does not remove the underlying mismatch between promised slash coverage and available collateral.  Our additive loss measure deliberately records obligations before recovery.  Extending the planner to value service-specific shortfalls would permit direct comparison of alternative recovery and insurance designs.

\paragraph{Scope.}
The model is deliberately a model of allocation under stochastic operational risk, not a complete model of protocol security.  It treats rewards, marginal slash probabilities, and common-factor exposures as inputs, and it records gross claims before any recovery rule is applied.  These choices make the separate roles of architecture, dependence, and competition transparent.  They also determine how the findings should be used: as comparative statics and stress-test mechanisms rather than forecasts of current returns or incident frequencies.  Endogenous attacks, operator effort, and post-slash recovery are natural extensions.

\section{Conclusion}
\label{sec:conclusion}

Correlated slashing must be analyzed jointly with collateral architecture.  The identity $\sigma^2G^2[\rho+(1-\rho)H]$ separates the nondiversifiable effect of common failures from the effect of collateral reuse on gross exposure.  This distinction explains the correlation floor under unique stake, leverage-like growth under shared collateral, and the reversal in how risk sensitivity affects breadth.

Strategic interaction adds a separate margin.  In proportional reward markets, individual allocation is strictly concave and the multi-allocator game has a unique, computable equilibrium, yet competition can still create excessive exposure and common-shock risk.  The framework turns architecture, dependence, and reward design into explicit economic objects that protocols can measure, stress test, and manage separately.

\section*{Use of generative AI}
The authors used OpenAI's ChatGPT and Codex to assist with theoretical exploration, numerical checks, code development, and manuscript editing.  All AI-assisted calculations, code, arguments, and prose were reviewed and validated by the authors, who take full responsibility for the methodology, results, and final manuscript.

\bibliographystyle{ACM-Reference-Format}
\bibliography{refs}

\appendix
\section{Reproducibility and Supplementary Results}
\label{app:reproducibility}

\subsection{Implementation and provenance}
The accompanying artifact separates environment generation, Gaussian-threshold risk, payoff and gradient evaluation, capped-simplex projection, equilibrium solvers, diagnostics, and reporting.  Its numerical core implements capped-simplex projection, analytical proportional-reward gradients, simultaneous updates, and projected-residual logic.  The paper-aligned implementation adds attainable Bernoulli covariance and forms the loss covariance
\[
V=D_\alpha\Cov(F)D_\alpha.
\]
It also supports fractional reuse caps and optional participation.  Separate modules compute best-response certificates, planner solutions, and alternative risk objectives.

Experiments A and D use the adaptive simultaneous projected method as their primary equilibrium engine.  Cyclic best response is an independent certificate and fallback; no fallback is used in the reported architecture or competition sweeps.  Each equilibrium is evaluated using
\[
\epsilon_{\mathrm{BR}}(x)
=
\max_i\left[\max_{y_i\in\cX_i}u_i(y_i,x_{-i})-u_i(x)\right]
\]
and
\[
\epsilon_{\mathrm{VI}}(x;\tau)
=
\norm{x-\Pi_{\cX}(x-\tau\cF(x))}_2.
\]
All plots and tables are generated from saved CSV files.  From a clean checkout, run
\begin{center}
\texttt{python scripts/reproduce\_all.py}
\end{center}
The command clears generated outputs, runs tests and Experiments A--F, rebuilds all reported artifacts, records software metadata, and writes a SHA-256 manifest.

\subsection{Synthetic parameterization}

Unless an experiment states otherwise, environments independently draw
$W_i\sim U[0.85,1.15]$, $R_j\sim U[0.105,0.160]$,
$b_j\sim U[1.15,2.00]$, $p_j\sim U[0.025,0.070]$,
$\alpha_j\sim U[0.35,0.80]$, and $\eta_i\sim U[1.1,1.9]$.
These are normalized payoff units, not annualized forecasts.

The factor vector contains one global factor, four operator factors, three software-client factors, and four cloud factors.  Each position loads on the global factor and on exactly one factor from each of the other three families.  For a one-based position index $j$, define
\[
o(j)=1+((j-1)\bmod 4),
\qquad
c(j)=1+\left(\left\lfloor\frac{j-1}{4}\right\rfloor\bmod 3\right),
\]
and
\[
h(j)
=
1+\left(
2(j-1)+\left\lfloor\frac{j-1}{4}\right\rfloor
\right)\bmod 4.
\]
Here $o(j)$, $c(j)$, and $h(j)$ identify position $j$'s operator, client, and cloud groups.  The staggered assignments create overlapping rather than nested dependency clusters.

Let $(\omega_g,\omega_o,\omega_c,\omega_h)$ denote the variance shares assigned to the global, operator, client, and cloud components.  With all factor variables and idiosyncratic shocks mutually independent and standard normal, the latent variable for position $j$ is
\[
\begin{aligned}
Y_j={}&
\sqrt{\omega_g}Z_g
+\sqrt{\omega_o}Z^{\mathrm{op}}_{o(j)}
+\sqrt{\omega_c}Z^{\mathrm{client}}_{c(j)}
+\sqrt{\omega_h}Z^{\mathrm{cloud}}_{h(j)}\\
&+
\sqrt{1-\omega_g-\omega_o-\omega_c-\omega_h}\,\varepsilon_j.
\end{aligned}
\]
Consequently, for distinct positions $j$ and $k$, their latent correlation is
\[
\beta_j^{\trans}\beta_k
=
\omega_g
+\omega_o\1\{o(j)=o(k)\}
+\omega_c\1\{c(j)=c(k)\}
+\omega_h\1\{h(j)=h(k)\}.
\]
Thus every pair shares the global component, while additional dependence arises when the positions share an operator, software client, or cloud provider.

\begin{table}[H]
\centering
\small
\begin{tabular}{@{}lccccc@{}}
\toprule
Regime & Global & Operator & Client & Cloud & Idiosyncratic \\
\midrule
Near independent
& $0.002$ & $0.008$ & $0.004$ & $0.004$ & $0.982$ \\
Clustered
& $0.030$ & $0.280$ & $0.100$ & $0.080$ & $0.510$ \\
Global plus clustered
& $0.380$ & $0.100$ & $0.040$ & $0.030$ & $0.450$ \\
\bottomrule
\end{tabular}
\caption{Variance shares in the recurring latent-factor regimes.}
\label{tab:factor-regimes}
\end{table}

These are shares of the unit variance of the latent Gaussian variable, not pairwise correlations between the binary slash indicators.  Thresholding at $\Phi^{-1}(p_j)$ preserves each position's marginal slash probability and maps the latent correlations into the coherent Bernoulli covariance matrix in \eqref{eq:binary-cov}.  The near-independent regime is almost entirely idiosyncratic, the clustered regime emphasizes shared operational dependencies, and the global-plus-clustered regime gives every pair of positions a substantial common component.  These regimes are controlled stress-test designs rather than empirical estimates, and their factor topology is held fixed across random seeds.

Two robustness experiments use related parameterizations.  Experiment C fixes total systematic variance at $0.55$, varies the global share over $\{0,0.10,0.25,0.40,0.55\}$, and divides the remainder among operator, client, and cloud factors in proportions $0.50$, $0.27$, and $0.23$.  The stress environments in Experiment E use shares $(0.58,0.18,0.08,0.05)$, leaving $0.11$ of latent variance idiosyncratic.

\begin{table}[H]
\caption{Complete computational design.  Seeds are fixed in the artifact; the reproduction workflow generates the processed outputs.}
\label{tab:appendix-experiment-design}
\centering
\small
\begin{tabularx}{\linewidth}{@{}p{.06\linewidth}p{.39\linewidth}X@{}}
\toprule
& Main sweep & Scale and evaluation \\
\midrule
A & $\kappa=1,\ldots,12$; three regimes & $n=4$, $m=12$, five seeds, $80{,}000$ draws; 180 equilibria \\
B & True, diagonal, and rank-three covariance & $m=16$, ten seeds, $120{,}000$ draws; $250$--$5{,}000$ estimation samples \\
C & Global factor share $0$--$0.55$; $\kappa\in\{1,4\}$ & $m=16$, eight seeds, $100{,}000$ draws; 400 portfolios \\
D & $n=1,2,4,8,12$; three background scales & $m=8$, total capital $4$, $\kappa=2$, five seeds; 75 equilibria \\
E & Three conditioning regimes and four sizes & Two seeds, 24 environments; 156 solver runs \\
F & Entropic $\theta=0.02$--$2$; CVaR $\zeta=0.05$--$1$ & $m=12$, $\kappa=4$, five seeds; $30{,}000$ train/$180{,}000$ test \\
\bottomrule
\end{tabularx}
\end{table}

\subsection{Validation}
The Gaussian-threshold covariance, analytical loss moments, gradients, projection, symmetric equilibrium, and cross-solver agreement are tested numerically.  The higher-sample validation suite reports the following errors.
\begin{table}[H]
\centering
\small
\begin{tabular}{lrrc}
\toprule
Test & Observed error & Tolerance & Passed \\
\midrule
Factor covariance & $4.36\times10^{-4}$ & $1.50\times10^{-3}$ & yes \\
Loss mean & $2.36\times10^{-4}$ & $2.00\times10^{-3}$ & yes \\
Loss variance & $4.17\times10^{-5}$ & $2.00\times10^{-3}$ & yes \\
Equilibrium exploitability & $6.03\times10^{-10}$ & $1.00\times10^{-7}$ & yes \\
Equilibrium VI residual & $1.14\times10^{-6}$ & $2.00\times10^{-6}$ & yes \\
Cross-solver $\ell_2$ spread & $3.03\times10^{-4}$ & $5.00\times10^{-4}$ & yes \\
\bottomrule
\end{tabular}

\caption{Numerical validation of the implementation.}
\label{tab:validation}
\end{table}

\subsection{Solver robustness}
The certified fixed-step method follows directly from \Cref{thm:equilibrium}, but its worst-case step bound is conservative.  The artifact therefore compares it with adaptive projected updates and three best-response orders.  Best-response dynamics converge in all of our experiments, although this behavior is not covered by our theoretical convergence result. \Cref{tab:solver-robustness} reports the aggregate diagnostics across the 24 solver environments.
\begin{table}[H]
\caption{Selected solver diagnostics.  Runtime is machine dependent; exploitability and VI residual are the substantive checks.}
\label{tab:solver-robustness}
\centering
\small
\begin{tabular}{@{}lrrr@{}}
\toprule
Method & Reached Tol. & Median iter. & Max VI residual \\
\midrule
Cyclic best response & $100\%$ & $7$ & $2.3\times10^{-6}$ \\
Adaptive projected & $100\%$ & $1{,}498$ & $4.6\times10^{-9}$ \\
Fixed-step projected & $25\%$ & $30{,}000$ & $1.6\times10^{-4}$ \\
Baseline $2/\sqrt t$ schedule & $46\%$ & $30{,}000$ & $1.7\times10^{-5}$ \\
\bottomrule
\end{tabular}
\end{table}

\subsection{Additional robustness}

We also test whether the allocation results depend on the quadratic mean--variance penalty.  Mean--variance and exact entropic allocations nearly coincide at low sensitivity, as predicted by \eqref{eq:entropic-risk}.  At sensitivity $2$, entropic risk reduces gross exposure from $3.818$ to $3.093$ and $99\%$ CVaR from $1.594$ to $1.262$.  A separate objective that shifts weight from expected loss toward $95\%$ CVaR can reduce exposure further and make nonparticipation optimal. Together, these results trace how stronger sensitivity to tail losses reduces exposure and can ultimately make nonparticipation optimal.

\end{document}